
\input vanilla.sty
\scaletype{\magstep1}
\baselineskip 16pt

\define\nlb{\par\vskip 14truept}

\define\nls{\par\vskip  2truept}

\def\cl{\centerline}
\def\mo{$M_{\odot}$}

\parindent = 20truept
\predisplaypenalty =0
\abovedisplayskip = 3mm plus 6truept minus 4truept
\belowdisplayskip = 3mm plus 6truept minus 4truept
\abovedisplayshortskip = 0mm plus 6truept
\belowdisplayshortskip = 2mm plus 6truept minus 4truept
\def\abstract#1 {\par{\narrower\noindent {\bf Abstract.} \hskip 2mm #1\par}}
\def\figure#1#2 {\par{\narrower\noindent {\bf Figure #1.} \hskip 2mm #2\par}}
\def\table#1#2 {\par{\narrower\noindent {\bf Table #1.} \hskip 2mm #2\par}}
\def\longref#1 {\par{\hangindent=20pt \hangafter=1 #1 \par}}

\def\begref{\vskip1mm\begingroup\let\INS=N}
\def\ref{\goodbreak\if N\INS\let\INS=Y\vbox{\noindent{\tenbf
}}\fi\hangindent\parindent
\hangafter=1\noindent\ignorespaces}

\cl {\bf QUARK STARS IN CHIRAL COLOUR DIELECTRIC MODEL}
\par\vskip 1in
\cl {Sanjay K. Ghosh and Pradip K. Sahu}
\nls
\cl {Institute of Physics}
\nls
\cl {Bhubaneswar 751 005}
\nls
\cl {INDIA}
\par\vskip 1.8in
\noindent\cl{\bf Abstract}
\nls
    The nonlinear chiral extension of colour dielectric model has been
used in the present work to study the properties of quark stars.
Assuming that the square of meson fields develop nonzero expectation
value, the thermodynamic potential for charge neutral interacting two
and three flavour quark matter, in beta equilibrium, has been
calculated up to second order in quark gluon interaction term in the
Lagrangian. The equation of state has been found to be softer for
higher quark masses and interaction strength. The quark stars
properties are found to be dependent on EOS. The mass, radius, moment
of inertia and gravitational red shift decrease as the equation of
state becomes more soft.
\vfil
\nls
\eject
\cl {\bf Introduction}

The quark structure of hadrons suggests the possibility of a phase
transition from nuclear to quark matter at high density. Several
calculations have been done to investigate such phase transitions[1].
There are speculations about the existence of quark stars or
quark core in neutron stars.Calculations for such cases have
also been done using different models[2].
The present situation demands a clear cut way to distinguish between
neutron and quark stars. It is generally suggested that one should
use measurement of the surface gravitational red shift $z$, as it will
be different for different equations of state. Also a knowledge
of neutrino flux may help as the flux from quark matter expected to be
quite larger in comparison with neutron star matter[3]. In such a situation
one should really use more realistic models to understand the
possible characteristic of quark stars.
A number of QCD-motivated models have been developed and successfully
applied to  study  the properties of light baryons[4]. The colour
Dielectric Model has the added characteristic that here confinement
is generated dynamically through the interaction of scalar field
(colour dielectric
field) with quarks[5]. The chiral version of this model has been
prescribed
by Banerjee and et. al[6]. This model has also been used successfully for
baryon spectroscopy[7] as well as quark matter calculations[8-9]. In the
present paper, using the ansatz given in ref.[9], we have investigated
the properties of cold, charge neutral quark matter and applied it to study
quark star properties. We have compared our results with those
obtained from bag model, where meson fields are not included. We
have also studied the effect of quark-gluon interactions and
beta equilibration on these properties. The paper is organised as follows.
In section 2 we define the model. In section 3 we give the results and
discussions and finally the conclusions are given in section 4.
\nls
\cl {II Chiral colour dielectric model}

Following the usual three flavour version of cloudy bag models[10],
 the Lagrangian density of the  nonlinear  version  of  the
chiral CDM may be written as,

	$$\eqalignno{L(x)=& \bar\psi(x)\big \{ i\gamma^{\mu}\partial_{\mu}-
(m_{0}+(m/\chi(x)) U_{5}) + (1/2) g
\gamma_{\mu}\lambda_{a}A^{a}_{\mu}(x)\big \}\psi \cr
& +(f^{2}_{\pi}/4) Tr ( \partial_{\mu}U
\partial^{\mu}U \dagger ) -
(1/2)m^{2}_{\phi}\vec \phi^{2}(x)-(1/4)
\chi^{4}(x)(F^{a}_{\mu\nu}(x))^{2} \cr
&+ (1/2) \sigma^{2}_{v}(\partial_{\mu}\chi(x))^{2} - U(\chi) & (1)  \cr}$$

where $$\eqalignno{U & = e^{i\lambda_{a}\phi^{a}/f_\pi} & (2) \cr}$$ and
$$\eqalignno{U_{5} & =  e^{i\lambda_{a}\phi^{a}\gamma_{5}/f_{\pi}} & (3) \cr}$$
 $\psi(x)$, $A_{\mu}(x)$, $\chi(x)$ and $\phi(x)$ are quark,
gluon, scalar ( colour
dielectric ) and meson fields respectively, $ m_{\phi}$ and m  are the meson
and quark masses, $ f_{\pi}$ is the pion  decay  constant,
 $F_{\mu\nu}(x)$ is the
usual colour electromagnetic field tensor, g is the colour coupling
constant and $\lambda_{a}$ are the Gell-Mann matrices.
The flavour symmetry breaking is
incorporated in the Lagrangian through the quark mass term
$(m_{0}+m/\chi U_{5})$, where $m_{0}=diag(0,0,m_{0})$ and
$m=constant$. So u and d quark have equal masses and s quark mass
$m_{s}= m_{0} + m$.  The
self-interaction of the scalar field is assumed to be of the form,
    $$\eqalignno{ U(\chi) & = \alpha B \chi^2(x)[
1-2(1-2/\alpha)\chi(x)+(1-3/\alpha)\chi^2(x)] & (4) \cr} $$
so that $ U(\chi)$ has an absolute minimum at $ \chi=0$ and a
secondary
minimum at $\chi=1$. The interaction of the scalar  field  with  quark
and gluon fields is such that quarks and gluons cannot  exist  in
the region where $\chi= 0$. In  the  limit  of  vanishing  meson  mass,  the
Lagrangian of eq(1) is invariant under chiral transformations of quark
and meson fields. In our  calculation,  we  assume  that,  because  of
 nonvanishing  quark
and antiquark densities in the medium, the square of  the  meson  field,
$<\vec\phi^{2}>$ develops a nonzero expectation value[9]. We however
assume that the
expectation value of $\vec\phi$  vanishes in the  medium.
 Our assumption is  quite  similar  to
that used in  linear  chiral  CDM  calculations, that the $\sigma$ field
develops  nonzero  expectation value in the medium, although  the
expectation value of pion field is  zero. For an isospin symmetric
matter, this is a reasonable assumption since there is no preferable
direction in isospin space. For an infinite system of quarks,  we  can
assume  that  $<\vec\phi^{2}>$  is  independent  of space and time.
\nls
\cl{\bf Two flavour matter}
 In case of two flavour matter $m_{s}=0$, $\lambda_{a}$ becomes
usual Pauli matrices and the Lagrangian reduces to the form as
given in ref.[9].
  The  pion
excitations ( $\vec\pi^{'}$) are then defined in terms of
fluctuations about
$<\vec\pi^{2}>$, so that $\vec\pi^{2} = <\vec\pi^{2}>+
\vec\pi^{'}^{2}$ . Defining
 $F= <\vec\pi^{2}>/f^{2}$, the chiral CDM
Lagrangian can be written in terms of F and pion excitations
$\vec\pi^{'}$[9]. However in the present calculation
we have considered only quark-gluon interaction. Even if one includes
pion-quark interaction the results will remain the same as its
contribution is much less compared to quark-gluon interaction.
In the mean field  approximation, $\chi$
 is independent of space and time for uniform  quark  matter.  The
mean field values of $\chi$ and F are  determined  by  minimising  the
thermodynamic potential $\Omega$.  Equivalently,  one  could  solve  the
classical equations of motion for $\chi$  and  F  obtained  from  the
Lagrangian. Considering the equation of motion at zero
temperature we find effective quark mass $ m^{*}
=(m/\chi)(1-F/3)e^{-F/3} $    and $\epsilon_{\pi}(k)= (k^2 +
m_{\pi}^{*}^{2})^{1/2}$  , where
$m_{\pi}^{*} = m_{\pi} (3/(2+1.5 (1-e^{-2F/3})/F))^{1/2}$. Thus
effective pion mass increases for
nonzero F.
We calculate the equation of state for charge neutral quark matter
following the procedure described in ref.[2,11]. The condition of charge
neutrality is invoked through the relation $$ \eqalignno{ (2/3) n_{u}-
(1/3) n_{d}- n_{e} & = 0 & (5) \cr}$$
Beta equilibrium can be incorporated by putting an additional
restriction on chemical potential,
$$\eqalignno{\mu_{d} & = \mu_{u}+ \mu_{e} & (6) \cr}$$
where $\mu_{i}= (m^{*}^{2}+ k_{i}^2)^{1/2} $for i= u,d, $k_i$ being
Fermi momentum.
$n_{i}= - \partial{\Omega}/\partial\mu_{i}$ and baryon density
$n_{B}=(1/3)\sum_{i}(n_{i})$. We have taken electron mass to be zero so
that $\mu_{e}= k_{e}$ and $n_{e}= k_{e}^{3}/(3\pi^{2})$.
Using the above conditions we have calculated thermodynamical
potential $\Omega$ up to second order in quark-gluon interaction using
standard imaginary time formalism[12]. The
thermodynamic quantities are calculated from $\Omega$ using standard formulae.
Thus pressure $P(\mu)= -\Omega(\mu)$, energy density $E(\mu)=
\Omega(\mu)+ \sum_{i}\mu_i n_{i}$ where i runs over u, d and e.
\nls
\cl {\bf Three flavour matter}
The true ground state of hadrons may be strange matter, that is, the
matter consist of almost equal number of u, d and s quarks plus a
small number of electrons as conjectured by Witten[13].
We have found[19] that for three flavour matter only $<\vec\pi^{2}>$
develops non zero
value and $<K^2>$ = $<\eta^2>$= 0 which means that strange quark mass
remains constant in the
medium. The u and d quark masses change in the manner given in two
flavour case. The chemical equilibrium among the constituents implies
$$\eqalignno{\mu_{d} & = \mu_{u}+ \mu_{e} & (7)\cr}$$ and $$\eqalignno{
\mu_{s} & = \mu_{u}+\mu_{e} & (8) \cr}$$
Charge neutrality gives $$ \eqalignno{(2/3) n_{u} - (1/3) n_{d} -(1/3)
n_{s} - n_{e} & = 0 & (9) \cr} $$  Baryon density
$n_{B}=(1/3)\sum_{i}(n_{i})$ where i= u,d,s. With these conditions we
calculate thermodynamic potential up to second order in quark-gluon
interaction.
\nls
\cl{\bf Quark Star Structure}
\nls
 From the studies of quark matter[14] it is predicted that the mass (M) of
quark stars $\simeq$ \mo  ( \mo  solar mass ) and radius (R)
$\simeq 10$ km. These so-called quark stars have rather different
mass-radius relationship than neutron stars, but for stars of mass
$\simeq 1.4 $\mo , the structure parameters of quark stars are
very similar to those of neutron stars.
\nls
The mass and radius for non-rotating quark stars are
obtained by integrating the structure equations of a relativistic
spherical static star composed of a
perfect fluid which is derived from
Einstein's equation by Oppenheimer and
Volkoff[15-16].
$$\eqalignno{\frac{dP}{dr} &= - \frac{G (\rho + P/c^2) (m + 4\pi r^3 P/c^2)}
{r^2 (1-2Gm/rc^2)} & (10) \cr
 \frac{dm}{dr} &= 4\pi r^2\rho & (11) \cr}$$
 where G is gravitational constant, c is velocity of light, P is the
pressure and $\rho$ is the density.
\noindent The potential  function, $\nu(r)$, relating the element of
proper time
to the element of time at $r=\infty$ is given by
$$\eqalignno {\frac{d\nu}{dr} &= \frac{2G}{r^2}\frac{(m + 4\pi r^3
P/c^2)}{(1-2Gm/rc^2)} & (12) \cr}$$
with boundary condition, $\nu(R) = ln(1-{2GM/R c^2})$.
Then the moment of inertia is given by
$$\eqalignno{\frac{dI}{dr} &= \frac{8\pi}{3}\frac{(\rho +
P/c^2)}{(1-2Gm/rc^2)^{1/2}} e^{-\nu(r)/2}\frac{\Omega -
\omega(r)}{\Omega} & (13) \cr}$$
where $\omega(r)$ is the angular velocity of the local inertial
frame. Since for slow rotation it is much smaller than the star's angular
velocity, $\Omega$, we neglect it and eqn.(13) can be written as
$$\eqalignno{\frac{dI}{dr} &= \frac{8\pi}{3}\frac{(\rho +
P/c^2)}{(1-2Gm/rc^2)^{1/2}} e^{-\nu(r)/2} & (14) \cr}$$

\noindent For a given EOS, $P(\rho)$, and a given central density,
$\rho(r=0) = \rho_c $, the Eqs. (10-14) are integrated numerically with
the boundary condition:
$$\eqalignno{ m(r = 0) &= 0 & (15) \cr}$$
\noindent to give R and M.  The radius R is defined by the point
where P $\simeq$ 0.
 The total gravitational
mass , moment of inertia and gravitational red shift $z$ are  then given by:
$$\eqalignno{ M = m(R)., I = I(R), z=& ( 1- 2GM/Rc^2 )^{-1/2}- 1
) & (16) \cr}$$
\nls
\cl {\bf Results and Discussion}
\nls

The equation of state (EOS) and the properties of quark star are calculated
for a number of parameter set of chiral CDM (B, m, $\alpha$ and g). For two
flavour sector these are obtained by fitting nucleon and delta masses.
For three flavour sector we take  strange quark mass in the
range 150-400 MeV ($m_0=$ 24.1 to 274.1 MeV) along with two flavour
parameter sets. Here we have
given the results for a representative parameter set ($B^{1/4}=
157.4$
MeV, m(u and d)= 125.9 MeV,$ m_{s}= 250$. MeV, $\alpha= 36.$, g=
1.173 ) for
a detailed discussion. We find that increment in  quark masses
and quark-gluon coupling makes the EOS softer. The mean field value
of $\chi$ remains close to 1 at different densities for all parameter
sets. Therefore the pressure due to scalar field $\simeq -U(\chi=1
)\simeq -B$ and the net pressure vanishes when the pressure due to
matter fields$\simeq B$. So the increase  in B increases the density
at which the pressure of the system vanishes. Below this density the
quark matter system can not exist. For the above parameter set this
density lies between 2-3 times nuclear matter density for both two
and three flavour matter. As mentioned earlier, we have
calculated $\Omega$ up to second order in g. The EOS for two and three
flavour neutral quark matter, with and without gluon interaction are
shown in fig.1, where (a) and (b) are the two flavour sector without
and with interaction.The curves (c) and (d) corresponds to those for three
flavour sector in the same order.
For comparison we have also given simple massless quarks and
noninteracting MIT bag model EOS with bag
pressure $56 MeV/fm^{3}$ ( curve (e) ).
In table I and II we have given quarks and electron densities at
different baryon densities for two and three flavour matter
respectively.
\nls
In our calculation we have  $<\vec \pi^{2}>$ is non zero where
as $<K^{2}>$ and $<\eta^{2}>$ are zero. Since strange sector does
not couple with $\vec\pi$, only u and d quark masses will change with
density. Also we
have assumed $<\vec\pi^{+}^{2}>$, $<\vec\pi^{-}^{2}>$ and
$<\vec\pi^{0}^{2}>$ to
be same, so that u and d quarks have equal masses for all densities.
The variation of $m$ and $<\vec\pi^{2}>$ with density are
shown in fig.2. It is clear from the graph that u and d quarks
mass reduce to almost one third of its original value, even for
densities at which pressure goes to
zero. Also it increases due to
the presence of interaction.  The implication of such variation in
quark masses are discussed in detailes in ref.[9].
\nls
The results for quark star masses and radii are shown in figs. 3 and
4. The important regime of the EOS involves energy densities between
about $3.0\times 10^{15}$ to $4.5\times 10^{15} g/cm^{3}$. As far as
neutron stars are concerned, any acceptable EOS should predict
maximum mass at least equal to the masses of observed neutron star which
are 1.4 solar mass. But it is possible that there
exist quark stars with lower masses and radii. In fact there can be
quark cores inside neutron stars as well. It is evident from fig.3
that softening of EOS reduces the $M_{max}$, which suggests that the
attractive nature of quark-gluon interaction in matter reduces the
mass. Also the three flavour sector has lower $M_{max}$ compared to
two flavour sector. The $M_{max}$ goes down from 1.6 \mo  for
noninteracting two flavour to 1.26 \mo  for interacting three
flavour matter. The corresponding $R(M_{max})$ changes from 8.75 km to
7.33 km respectively. The mass M increases with radii unlike
neutron star case where mass decreases with increase in radius in much
of the range so that there is a minimum mass.
\nls
The simple MIT bag model EOS ( i.e. all quark masses equal to
zero, and the coupling constant goes to zero ) can be written
analytically
$$\eqalignno{ P = & (1/3)(\epsilon - 4 B) & (17) \cr}$$
where $\epsilon$ is the total energy density. The eqn.(17) is
independent of the number of quark flavours. This model does not
contain meson fields at all. Whereas the present model contains
meson fields as well as the quark messes evolve dynamically with
densities. The structure parameters of quark stars for simple
MIT bag model with bag pressure $56 MeV/fm^{3}$ are calculated
by integrating the structure eqns.(10-16). We found that the
$M_{max} = 1.97$ \mo, $R(M_{max})=10.8 $ km, $ I(M_{max})=2.20
\times 10^{45} g. cm^{2}$ and $z(M_{max})=0.47$, which are
higher than the model considered here for both two and three
flavour matter without and with interactions, see table III. In
comparison to extended bag model, the currently popular strange
star model[21], where the EOS is short-range quark-gluon
interactions perturbativley to second order in the coupling
constant $\alpha_{c}$, and the long-range interactions are taken
into account phenomenologically by the bag pressure term (B),
the structure parameters of our model are still less and the EOS
are soft.
\nls
Fig.4 shows the variation of mass with central density $\rho_{c}$.
Near pressure going to zero, mass rises very fast indicating the
irrelevance of gravity in this region. With rise in $\rho_{c}$, $M$/\mo
 reaches a maximum which corresponds to maximum mass a stable star can
have.
\nls
One observable in the study of pulsars is the moment of inertia, $I$ of the
star and gravitational red shift $z$. We plot $I$ versus $M$/\mo  and $z$
versus$M$/\mo  in fig.5 and 6 respectively. $I$ and $z$ both are found to
be lower for
softer EOS. Again bag model gives the maximum value.
The values of $R(M_{max})$, $M_{max}$, $I(M_{max})$ and
$z(M_{max})$ for the curves (a), (b), (c), (d) and (e) are given in
Table III.
\nls
\cl{\bf Conclusion}
We have described the properties of quark star and the effect of
gluon interaction on them in the framework of chiral colour
dielectric model. In the present calculation we find that the EOS
becomes softer with the inclusion of strange quarks as well as gluon
interaction. Thus the maximum mass, radius, moment of inertia and
surface red shift decrease with increase in s quark mass and gluon
interaction. Observationally, the masses of neutron stars are
estimated from compact binary systems, one member of which is
pulsar. Most precise estimate comes from PSR 1913+16, which gives
1.442 $\pm$ 0.003 \mo[17]. The recent measurement for vela x- 1 pulsar
gives the maximum mass to be 1.77 $\pm$ 0.21 \mo[18].
 Most of the field theoretical calculations for
neutron star mass predicts higher values[20].
The results of present calculation are compatible with the
observationally inferred values as one would expect the masses for
quark stars or quark core in neutron stars to be lower than the
 neutron stars.
So our results for overall structured reported here reveals no
clear differences between the  quark stars or quark cores
in neutron stars with the neutron stars. An additional knowledge
of neutrino emissivity will help to distinguish between the
neutron stars and quark stars which will be reported in a later work.
\nlb
\noindent The authors would like to thank S. C. Phatak and B. Datta
for useful discussions and comments.
\vfil
\eject
\cl{\bf Reference}
\item {1.} B.A.Freedman and L.D.MacLerran, Phys. Rev. D{\bf 17}
(1978)1169; J. I. Kapusta, Nucl. Phys. B{\bf 148}(1979)461
\item {2.} J. C. Collins and M. J. Perry, Phys. Rev. Lett. {\bf
34}(1975)1353; B. A. Freedman and L. D. MacLerran, Phys. Rev. D{\bf
17}(1978)1109; B. D. Serot and H. Uechi, Ann. of Phys. {\bf
179}(1987)272
\item {3.} B. Datta and et.al., Mod. Phys. Lett. A{\bf 3}(1988)1385;
A.Goyal and J.D.Anand,\hfil\break
 Phys.Rev. D{\bf 42} (1990) 992
\item {4.} R. Friedberg and T. D. Lee, Phys. Rev. D{\bf 25}(1982)1951;
E. Witten, Nucl. Phys. B{\bf 223}\hfil\break (1983) 422; A. P.
Balchandran and et.
al., Phys. Rev. D{\bf 27} (1983) 1153
\item {5.} H. B. Nielson and A. Patkos, Nucl. Phys. B{\bf 197}(1982)139;
A. Schuh and H. J. Pirner,\hfil\break  Phys. Lett. B{\bf 173} (1986) 19;
N. Aoki and
H. Hyuga, Nucl. Phys. A \hfil\break {\bf 505} (1989) 525; S. C. Phatak,
Phys. Rev. C{\bf
44} (1991) 875
\item {6.} M. K. Banerjee and et. al. in Chiral Solitons, Eds. K. F. Liu
(World Scientific)1987
\item {7.} H. Kitagawa, Nucl. Phys. A{\bf 487}(1988)544; S. Sahu and
S. C. Phatak, Mod. Phys. Lett. A{\bf 7}(1992)709
\item {8.} W. Broniowski and et. al., Phys. Rev. D{\bf 41}(1990)285;
J. A. McGovern and et. al., J. of Phys. G{\bf 16}(1990)1561
\item {9.} Sanjay K. Ghosh and S. C. Phatak J. Phys. G {\bf 18}(1992)755
\item {10.} L. H. Chan in Chiral Solitons, Eds. K. F. Liu,(World
Scientific)1987; J. A. McGovern and M. C. Birse, Manchester Preprint,
MC/TH 91/1992
\item {11.} J. Ellis and et. al. Nucl. Phys. B{\bf 348}(1991)345
\item {12.} B. A. Freedman and L. D. McLerran, Phys. Rev. D{\bf
16}(1977)1130,1147,1169
\item {13.} E. Witten, Phys. Rev. D{\bf 30}(1984)272
\item {14.} E. Farhi and R. L. Jaffe, Phys. Rev. D{\bf 30}(1984)2379; D{\bf
32}(1985)2452; C. Alcock and et. al. Astrophys. J. {\bf 310}(1986)261;
P. Haensel and et. al., Astr. Astrophys.,{\bf 160}(1986)121
\item {15.} J. R. Oppenheimer and G. M. Volkoff, Phys. Rev. {\bf 55}(1939)374
\item {16.} C. W. Misner, K. S. Thorne and J. A. Wheeler, Gravitation
(Freeman, \hfil\break San Fransisco, 1970)
\item {17.} J. H. Taylor and J. M. Weisberg, Astrophys. J. {\bf 345}(1989)434
\item {18.} F. Nagase, Publ. Astr. Soc. Japan, {\bf 41}(1989)1
\item {19.} Sanjay K. Ghosh and S. C. Phatak, to be published.
\item {20.} R. C. Malone and et. al., Astrophys. J. {\bf
199}(1975) 741 and B. Arntsen and E. Ostgaard Phys. Rev. C. {\bf
30 }(1984) 335
\item {21.} B. Datta and et. al. Phys. Lett. B {\bf 283}(1992) 313
\eject
\cl {\bf Table.I}
\nlb
Different contributions to number density of two flavour matter,
without and with interaction;
$n_B=$, $n_u$, $n_d$, and $n_e$ are baryon, u quark, d quark and
electron number densities respectively.
\par\vskip 3in
\hskip .8in Noninteracting \hfil\hskip .2in  Interacting \hfil \nls
\hskip .6in number densities ($fm^-3$)\hskip .5in  number densities
($fm^-3$)\hfil\nlss
\midinsert$$\vbox{\offinterlineskip
\halign{&\vrule#&\strut\ #\ \cr
\noalign{\hrule}
height0pt&\omit&&\omit&&\omit&&\omit&&
\omit&&\omit&&\omit&&\omit&\cr
&\hfil{ }&&{ }&&{ }&&{ }&&{ }&&{ }&&{ }&&{ }&\cr
&\hfil$n_B$&&$n_u$&&$n_d$&&$n_e$&&$n_B$&&
$n_u$&&$n_d$&&$n_e$&\cr
&\hfil{ }&&{ }&&{ }&&{ }&&{ }&&{ }&&{ }&&{ }&\cr
\noalign{\hrule}
&\hfil{ }&&{ }&&{ }&&{ }&&{ }&&{ }&&{ }&&{ }&\cr
&\hfil 1.001&&1.006&&1.997&&5.6$\times
10^{-3}$&&1.035&&1.041&&2.064&&6.2$\times 10^{-3}$&\cr
&\hfil 0.801&&0.807&&1.598&&4.5$\times
10^{-3}$&&0.805&&0.809&&1.605&&4.8$\times 10^{-3}$&\cr
&\hfil 0.629&&0.633&&1.255&&3.5$\times
10^{-3}$&&0.613&&0.616&&1.222&&3.8$\times 10^{-3}$&\cr
&\hfil 0.482&&0.484&&0.961&&2.6$\times
10^{-3}$&&0.472&&0.475&&0.941&&2.8$\times 10^{-3}$&\cr
&\hfil 0.358&&0.360&&0.715&&1.9$\times
10^{-3}$&&0.352&&0.354&&0.702&&2.1$\times
10^{-3}$&\cr\noalign{\hrule} }}$$\endinsert
\vfil
\eject
\cl {\bf Table.II}
\nlb
Different contributions to number density of three flavour matter,
without and with interaction;
$n_B=$, $n_u$, $n_d$, $n_s$, and $n_e$ are baryon, u quark, d quark,
s quark and electron number densities respectively.
\par\vskip 3in
\hskip .8in Noninteracting \hfil\hskip .2in  Interacting \hfil \nls
\hskip .8in number densities ($fm^-3$)\hfil\hskip .2in  number
densities ($fm^-3$)\hfil\nlss
\midinsert$$\vbox{\offinterlineskip
\halign{&\vrule#&\strut\ #\ \cr
\noalign{\hrule}
height0pt&\omit&&\omit&&\omit&&\omit&&
\omit&&\omit&&\omit&&\omit&&\omit&&\omit&\cr
&\hfil{ }&&{ }&&{ }&&{ }&&{ }&&{ }&&{ }&&{ }&&{ }&&{ }&\cr
&\hfil$n_B$&&$n_u$&&$n_d$&&$n_s$&&$n_e$&&$n_B$&&
$n_u$&&$n_d$&&$n_s$&&$n_e$&\cr
&\hfil{ }&&{ }&&{ }&&{ }&&{ }&&{ }&&{ }&&{ }&&{ }&&{ }&\cr
\noalign{\hrule}
&\hfil{ }&&{ }&&{ }&&{ }&&{ }&&{ }&&{ }&&{ }&&{ }&&{ }&\cr
&\hfil 1.005&&1.005&&1.262&&0.748&&1.6$\times
10^{-4}$&&1.004&&1.004&&1.210&&0.798&&9.4$\times 10^{-5}$&\cr
&\hfil 0.804&&0.804&&1.040&&0.567&&1.9$\times
10^{-4}$&&0.808&&0.808&&1.001&&0.614&&1.2$\times 10^{-4}$&\cr
&\hfil 0.604&&0.604&&0.817&&0.390&&2.3$\times
10^{-4}$&&0.612&&0.612&&0.790&&0.434&&1.5$\times 10^{-4}$&\cr
&\hfil 0.483&&0.483&&0.679&&0.286&&2.7$\times
10^{-4}$&&0.471&&0.471&&0.634&&0.307&&1.9$\times 10^{-4}$&\cr
&\hfil 0.357&&0.357&&0.530&&0.183&&3.2$\times
10^{-4}$&&0.351&&0.351&&0.499&&0.203&&2.3$\times 10^{-4}$&
\cr\noalign{\hrule} }}$$\endinsert
\vfil
\eject
\centerline {\bf Table.III}
\nls\nls
The non-rotating quark star characteristics corresponding to $M_{max}$.
\midinsert$$\vbox{\offinterlineskip
\halign{&\vrule#&\strut\ #\ \cr
\noalign{\hrule}
height0pt&\omit&&\omit&&\omit&&\omit&&
\omit&\cr
&\hfil{ }&&{ }&&{ }&&{ }&&{ }&\cr
&\hfil Curves &&$M_{max}$(\mo)&&$R(M_{max})(km)$&&$I(M_{max})(\times
10^{+45} g \dot cm^2)$&&$z(M_{max})$&\cr
&\hfil{ }&&{ }&&{ }&&{ }&&{ }&\cr
\noalign{\hrule}
&\hfil{ }&&{ }&&{ }&&{ }&&{ }&\cr
&\hfil (a)&&1.60&&8.75&&1.20&&0.47&\cr
&\hfil (b)&&1.45&&8.30&&0.93&&0.44&\cr
&\hfil (c)&&1.40&&7.90&&0.82&&0.44&\cr
&\hfil (d)&&1.26&&7.33&&0.61&&0.42&\cr
&\hfil (e)&&1.97&&10.8&&2.20&&0.47&\cr\noalign{\hrule} }}$$\endinsert
\vfil
\eject
\cl {\bf Figure Caption}
\nlb
\item {Fig.1.} Plot for pressure P vs energy density E:
(a) Two flavour matter (TFM) without interaction, (b) TFM with
interaction, (c) Three flavour matter (ThFM) without interaction, (d)
ThFM with interaction and (e) Bag model EOS
\item {Fig.2.} Plot for $<\vec \pi^{2}>/f^2$ and $m^*$, the effective
quark mass vs baryon density $n_B$
\item {Fig.3.} Plot for $M$/\mo vs radius R
\item {Fig.4.} Plot for $M$/\mo vs central density $\rho_c$
\item {Fig.5.} Plot for moment of inertia $I$ vs $M$/\mo
\item {Fig.6.} Plot for gravitational red shift z vs $M$/\mo

\end